\documentclass[%
 reprint,
 amsmath,amssymb,
 aps,
]{revtex4-2}

\usepackage{graphicx}
\usepackage{dcolumn}
\usepackage{bm}
\usepackage{lineno}
\usepackage{tikz}
\usepackage{tikz-feynman}
\usepackage{caption}
\usepackage{amsmath}
\usepackage{float}
\usepackage{subcaption}
\usepackage{xcolor}

\begin{document}
\raggedbottom

\title{Long lived MeV scale bound state dark matter alleviates the ${}^7 Li$ tension}

\author{Pa. Gokhula Prasad}
\email{pa.gokhul1998@gmail.com}
\affiliation{Department of Physics, National Institute of Technology, Tiruchirappalli, India}

\author{V. Suryanarayana Mummidi}
\email{venkata@nitt.edu}
\affiliation{Department of Physics, National Institute of Technology, Tiruchirappalli, India}

\begin{abstract}
We investigate the impact of MeV scale unstable bound state dark matter on both thermal and non thermal Big Bang nucleosynthesis. Focusing on the weakly self interacting regime, we find that this framework can reproduce the theoretically predicted abundances through both short  and long lived bound states. While long lived bound states can alleviate the cosmological ${}^7$Li problem, they simultaneously lead to discrepancies with other light element abundances. We show that this tension is resolved by incorporating non thermal nucleosynthesis, which act on these non standard initial light element abundances to alleviate the ${}^7 Li$ tension without violating observational constraints on the other elements. Remarkably, across the range of $\widetilde{\rho}_B$ and $m_\chi$ considered, long lived bound states with lifetimes $10^0\,{\rm s}\lesssim\tau\lesssim10^2\,{\rm s}$ can simultaneously resolve the ${}^7$Li problem and a large parameter space remain consistent with the observed abundances of the other light elements. Our results demonstrate that non thermal effects are essential for obtaining a consistent  nucleosynthesis history in unstable bound state dark matter scenarios.  

\end{abstract}

\maketitle

\section{Introduction}
Primordial nucleosynthesis provides a crucial window into the early universe, serving as a powerful probe for Beyond Standard Model (BSM) physics. While Standard Big Bang Nucleosynthesis (SBBN) successfully predicts the primordial abundances of light elements such as  $^2\text{H}$ and  $^4\text{He}$ with remarkable precision, it suffers from a  discrepancy known as the cosmological lithium problem. The observed abundance of primordial  $^7\text{Li}$ inferred from metal poor stars is  three times lower than theoretical predictions from SBBN \cite{lp1,lp2,lp3,lp4,lp5}. 

To resolve this, numerous mechanisms have been explored, ranging from astrophysical solutions involving stellar depletion processes \cite{Korn} to BSM scenarios \cite{ex1,ex2,ex3,ex4,ex5,ex6,ex7} such as  gravitino decay\cite{BSM1,BSM2,BSM3}. However, non standard nucleosynthesis scenarios often encounter a severe constraint: mechanisms that successfully deplete lithium to observed levels frequently  overproduce or underproduce deuterium, thereby spoiling the agreement for other light elements \cite{d1,d2}.
In this Letter, we propose a BSM solution that evades these constraints by considering a self interacting complex scalar dark matter candidate, $\chi$, interacting via the exchange of a massless scalar mediator, $\Phi$. We extend this framework by accounting for the formation of unstable  bound states, which can decay either into $\Phi$ or Standard Model states.
We first compute the primordial light element abundances, incorporating the effects of the dark sector on background cosmology. Using these abundances as initial conditions, we study the impact of late time  photodisintegration and  find that this two stage process can resolve the primordial ${}^{7}\mathrm{Li}$  discrepancy while remaining consistent with observational constraints on the other light elements over a broad range of model parameter space.
This Letter is organized as follows: In Section~\ref{mod}, we briefly outline the model setup and the background cosmology. In Section~\ref{TH}, we compute the light element abundances resulting from primordial nucleosynthesis. In Section~\ref{nth}, we analyze how electromagnetic annihilation or decay products induce the photodisintegration of light elements, thereby constraining the model parameter space. Finally, Section~\ref{con} summarizes our study and presents our main conclusions.

\section{Model setup}
\label{mod}
In our framework, we consider a complex scalar dark matter  candidate  $\chi$, with a mass $m_\chi$ in the $\mathcal{O}(\mathrm{MeV})$. $\chi$ is self interacting through the exchange of a scalar mediator $\Phi$ . The relevant Lagrangian can be  written as
\begin{equation}
\mathcal{L}{\chi} \supset g_\chi \chi^*\Phi\chi
+ \mathcal{L}_{\rm portal}
+ \mathcal{L}_{\Phi},
\end{equation}
where $g_\chi$ denotes the strength of self interaction. In this work, we take $g_\chi \simeq 10^{-3}$ for rest of our analysis.

Since the bound state $\mathcal{B}$ is formed from a $\chi-\chi^*$ pair, it is unstable and can subsequently decay into pair of $\Phi$ or SM particles \cite{bo1}. Moreover, free DM particles can  self annihilate into a pair of mediators or into SM particles.

The presence of the additional dark degrees of freedom, the expansion of the Universe is modified through their contributions to the total energy density. Furthermore, the decay and annihilation of $\chi$ and $\mathcal{B}$ can inject entropy into the thermal bath, altering both the entropy density and the temperature evolution of the Universe. We discuss these effects on the background cosmology in detail in the Supplementary Material.
\begin{figure}
    \centering
    \includegraphics[width=\linewidth]{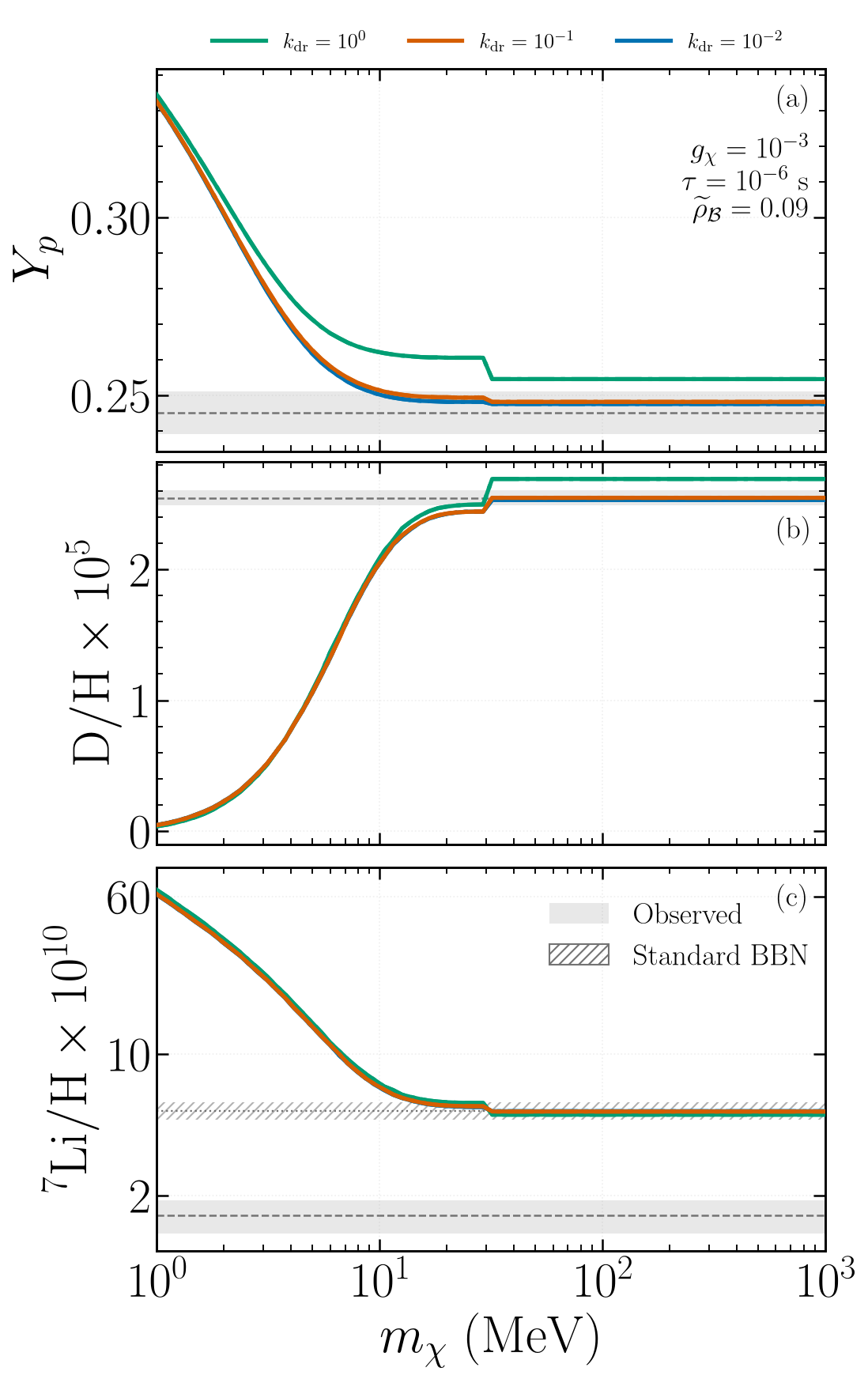}
\caption{Primordial light element abundances for  $m_\chi=1$--$10^3~\mathrm{MeV}$, varying $k_{\rm dr}$ and fixed $\widetilde{\rho}_{\mathcal{B}}=0.09$.}
\label{one}
\end{figure}
\begin{figure}
    \centering
     \makebox[\linewidth][c]{%
        \makebox[0.48\linewidth][c]{(a)}%
        \hfill
        \makebox[0.48\linewidth][c]{(b)}%
         
    }
    \includegraphics[width=\linewidth]{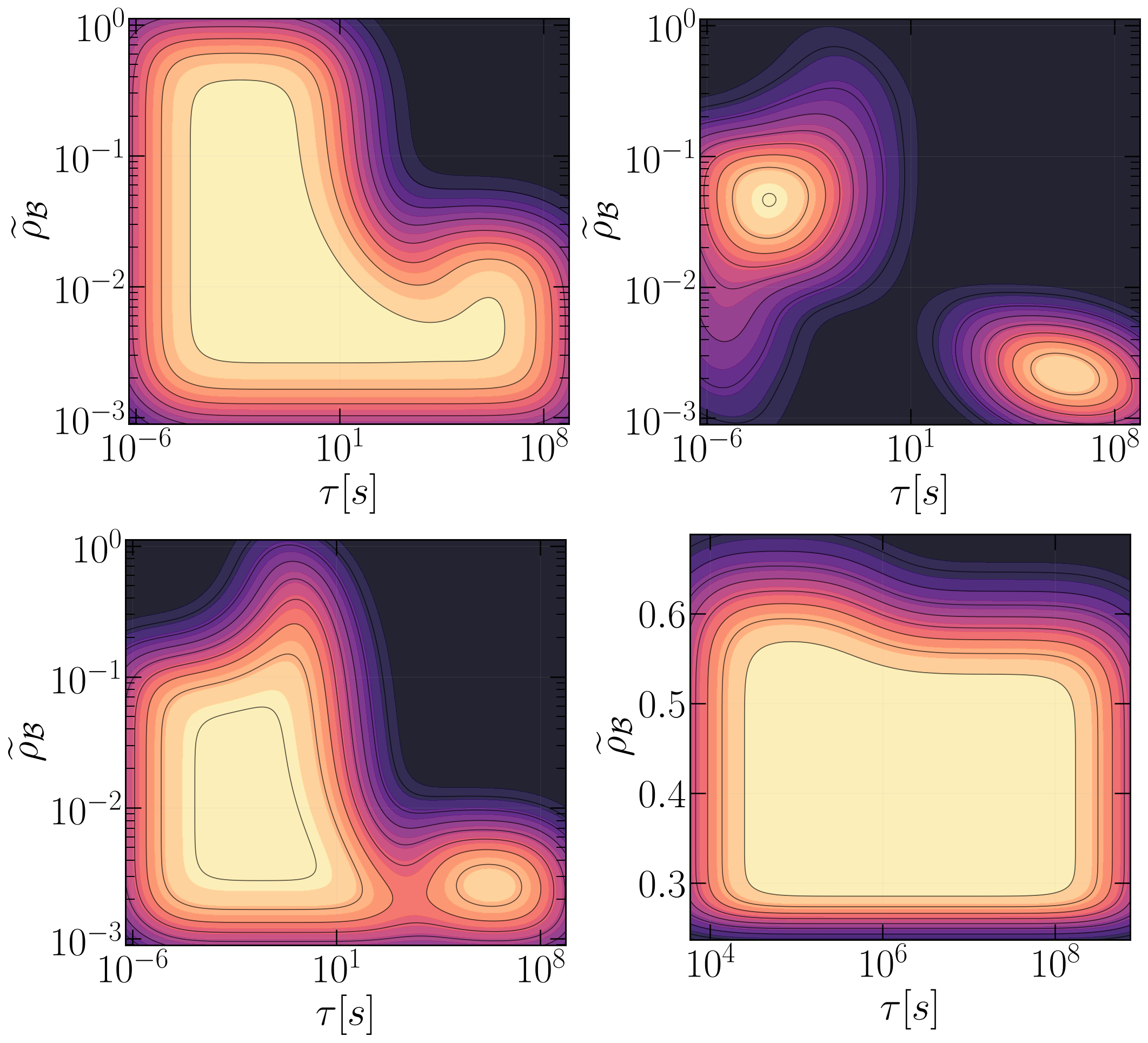}
     \makebox[\linewidth][c]{%
        \makebox[0.48\linewidth][c]{(c)}%
        \hfill
        \makebox[0.48\linewidth][c]{(d)}%
         
    }
\caption{KDEs of parameter points satisfying primordial abundance constraints: (a) $Y_p$, (b) ${\rm D/H}$, (c) theoretical ${}^7\mathrm{Li/H}$, and (d) observational ${}^7\mathrm{Li/H}$.}
\label{two}
\end{figure}
\begin{figure}
    \centering
    \includegraphics[width=\linewidth]{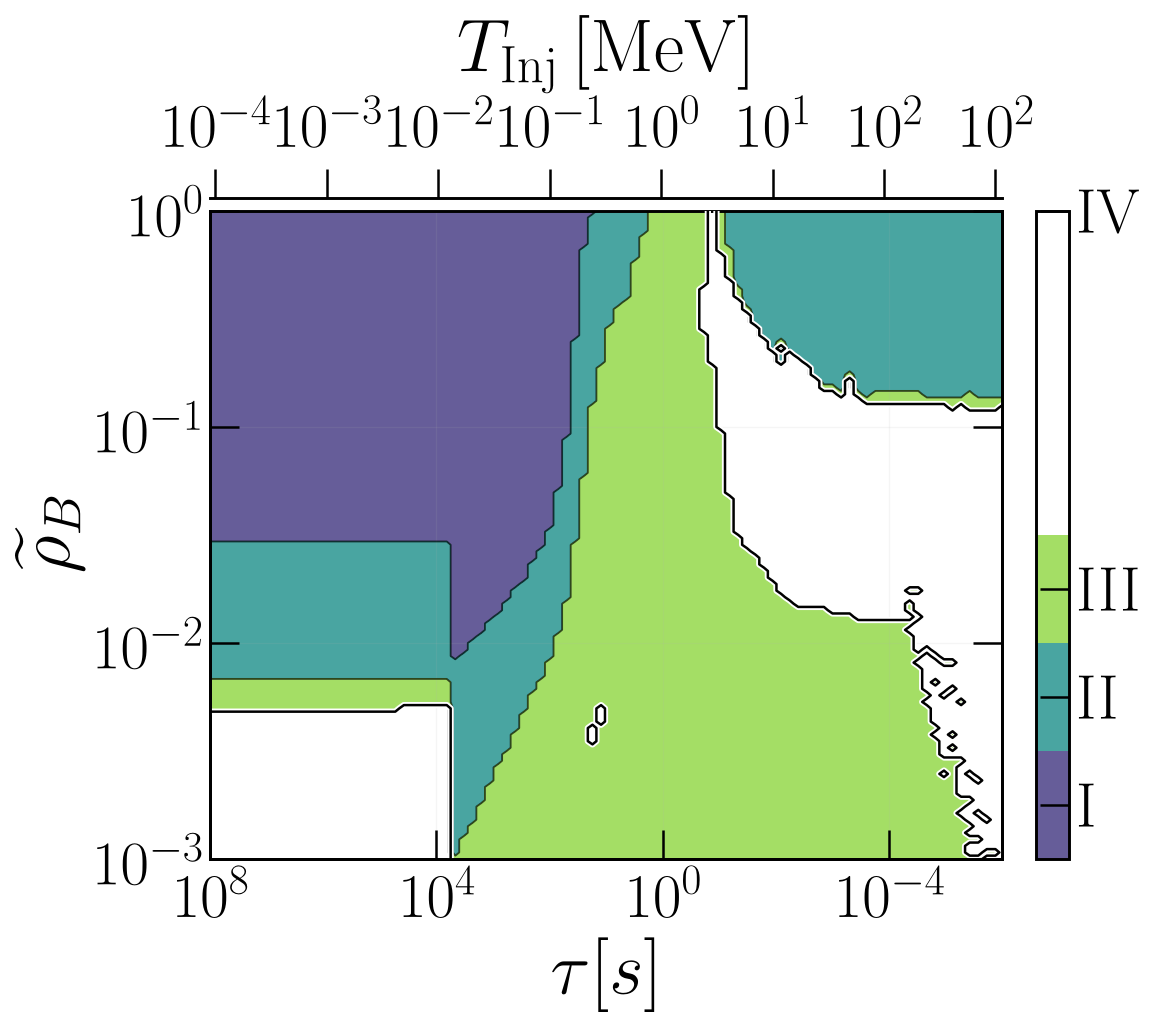}
\caption{Constrained parameter space from primordial light-element abundance bounds; the white region satisfies all constraints simultaneously.}
\label{three}
\end{figure}
In our previous work, Ref.\cite{me}, we demonstrated that this framework can successfully reproduce the observed dark matter relic abundance while remaining consistent with constraints on $N_{\rm eff}$. 
\section{Thermal Nucleosynthesis}
\label{TH}
We compute the primordial nuclear abundances in the presence of $\chi$, $\Phi$, and $\mathcal{B}$ by modifying \textit{AlterBBN V2.2} \cite{Alt} and  use the observational bounds at  2 $\sigma$ level  given in Table~\ref{tab:benchmarks} to constrain the parameter space. We take $m_\chi = 30~\mathrm{MeV}$ as the cutoff separating the two regimes. For $m_\chi > 30~\mathrm{MeV}$, dark matter freezes out before the onset of BBN, which begins at $T \simeq 1~\mathrm{MeV}$. In this regime, the bound states decay out of equilibrium during BBN, altering the evolution of both $\Phi$ and the SM thermal bath. Similar to \cite{bb1,bb2}, we assume that the SM decay products are predominantly electromagnetically interacting particles, with negligible decay into neutrinos, so that the injected energy primarily heats the electromagnetic plasma.

For $m_\chi < 30~\mathrm{MeV}$, dark matter freezes out during BBN. In this regime, we assume the bound state to be stable throughout nucleosynthesis and include its contribution to the evolution of the primordial abundances accordingly.

We adopt the following benchmark parameters:  lifetime of $\mathcal{B}$ before decaying  into SM particles $\tau=10^{-6}~\mathrm{s}$, bound state energy density $\widetilde{\rho}_{\mathcal B}\equiv\rho_{\mathcal B}/\rho_\gamma=0.09$ and $T_{Inj}$ is temperature of plasma the decay rate becomes comparable to the Hubble  rate. As shown in Fig.~\ref{one}, increasing $k_{\rm dr}=\frac{\rho_\Phi}{\rho_\gamma}|_{T=1 MeV}$ leads to an increase in  $Y_p$ and ${}^7\mathrm{Li}$ abundance, accompanied by a decrease in the deuterium abundance. This  can be understood from two complementary effects: First, the additional radiation component increases the Hubble rate, causing the weak interactions to freeze out earlier and thereby increasing the neutron to proton ratio. Second, the additional entropy contribution leads to an increase in the baryon to photon ratio $\eta$, which increases $Y_p$ and decreases the final ${\rm D/H}$ abundance \cite{Ko}.

For $m_\chi>30~\mathrm{MeV}$, as the bound states decay out of equilibrium, their energy density $\rho_{\mathcal B}$ decreases, reducing their contribution to the total energy density and consequently lowering the Hubble  rate and $\eta$ and this leads to a decrease in $Y_p$ and an increase in ${\rm D/H}$. 
In Fig.~\ref{two}, we perform a kernel density estimation  of the parameter points that satisfy the BBN abundance constraints in the $\rho_\mathcal{B}$--$\tau$ parameter space with $m_\chi = 700 MeV $. Panel (a) shows the KDE distribution of parameter points satisfying the helium abundance constraint. Although a large fraction of the parameter space is compatible with the helium constraint, regions with $\widetilde{\rho}_{\mathcal B}>10^{-2}$ and $\tau>10^{1}~s$ are largely excluded. As discussed in the Supplementary Material, these regions typically lead to an overproduction of ${}^4\mathrm{He}$. In contrast, parameter points with  $10^{-5}~s<\tau<10^{1}~s$ are more likely to satisfy the helium abundance constraint.

The allowed parameter space becomes substantially more restricted when the theoretical  deuterium and lithium-7  constraint is imposed, as shown in panel (b) and (c). Interestingly, two disconnected regions of parameter space satisfy the ${\rm \frac{{}^7 Li}{H}}$ constraint. For $10^{-3}<\widetilde{\rho}_{\mathcal B}<10^{0}$, the allowed points predominantly lie in the range $10^{-6}~s<\tau<10^{1}~s$, with the highest density of viable points occurring around $\tau\sim10^{-4}~s$. A second allowed region appears for $10^{-2}<\widetilde{\rho}_{\mathcal B}<10^{-3}$ and $10^{1}~s<\tau<10^{8}~s$, where the KDE peaks around $\tau \sim10^{6}~s$.  A  similar distribution is obtained for the parameter points satisfying the deuterium abundance bound.

Finally, panel (d) shows the KDE distribution of parameter points satisfying the observational ${}^7\mathrm{Li}$ abundance constraint. In this case, the allowed region is concentrated around $10^4~s<\tau<10^{8}~s$, but the corresponding  $\widetilde{\rho}_{\mathcal B}$ spans a comparatively narrow range, approximately $0.3\lesssim\widetilde{\rho}_{\mathcal B}\lesssim0.6$. 
\begin{figure}[t]
    \centering
     \makebox[\linewidth][c]{%
        \makebox[0.48\linewidth][c]{(a)}%
        \hfill
        \makebox[0.48\linewidth][c]{(b)}%
         
    }
    \includegraphics[width=\linewidth]{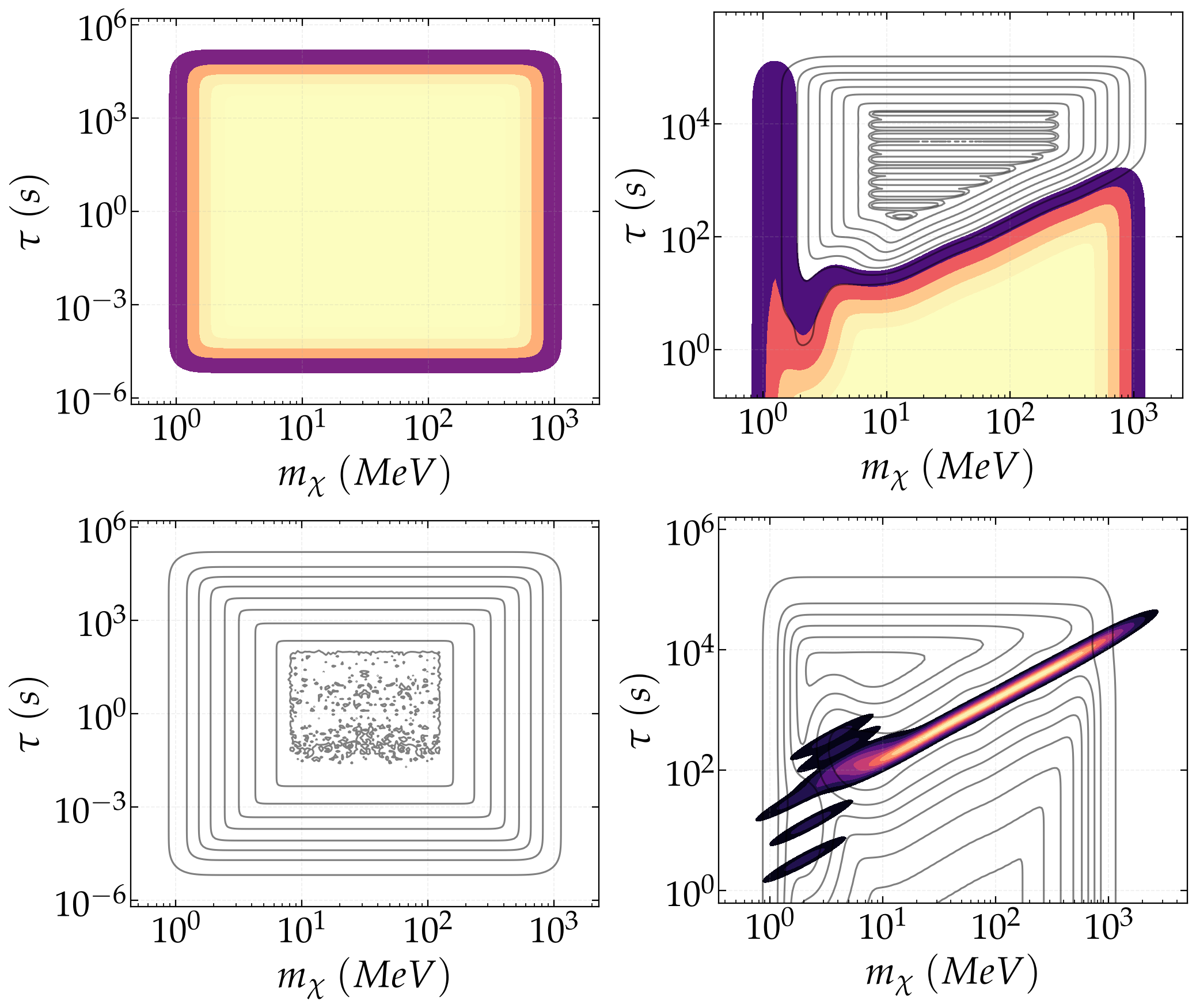}
     \makebox[\linewidth][c]{%
        \makebox[0.48\linewidth][c]{(c)}%
        \hfill
        \makebox[0.48\linewidth][c]{(d)}%
         
    }
\caption{KDE of parameter points satisfying the light element abundance constraints for inital SBBN abundance: (a) $Y_p$, (b) theoretical ${}^7\mathrm{Li/H}$, (c) ${\rm D/H}$, and (d) observational ${}^7\mathrm{Li/H}$. The black contour lines indicate regions that violate the corresponding observational bounds.}
\label{four}
\end{figure}
\begin{figure}[t]
    \centering
     \makebox[\linewidth][c]{%
        \makebox[0.48\linewidth][c]{(a)}%
        \hfill
        \makebox[0.48\linewidth][c]{(b)}%
         
    }
    \includegraphics[width=\linewidth]{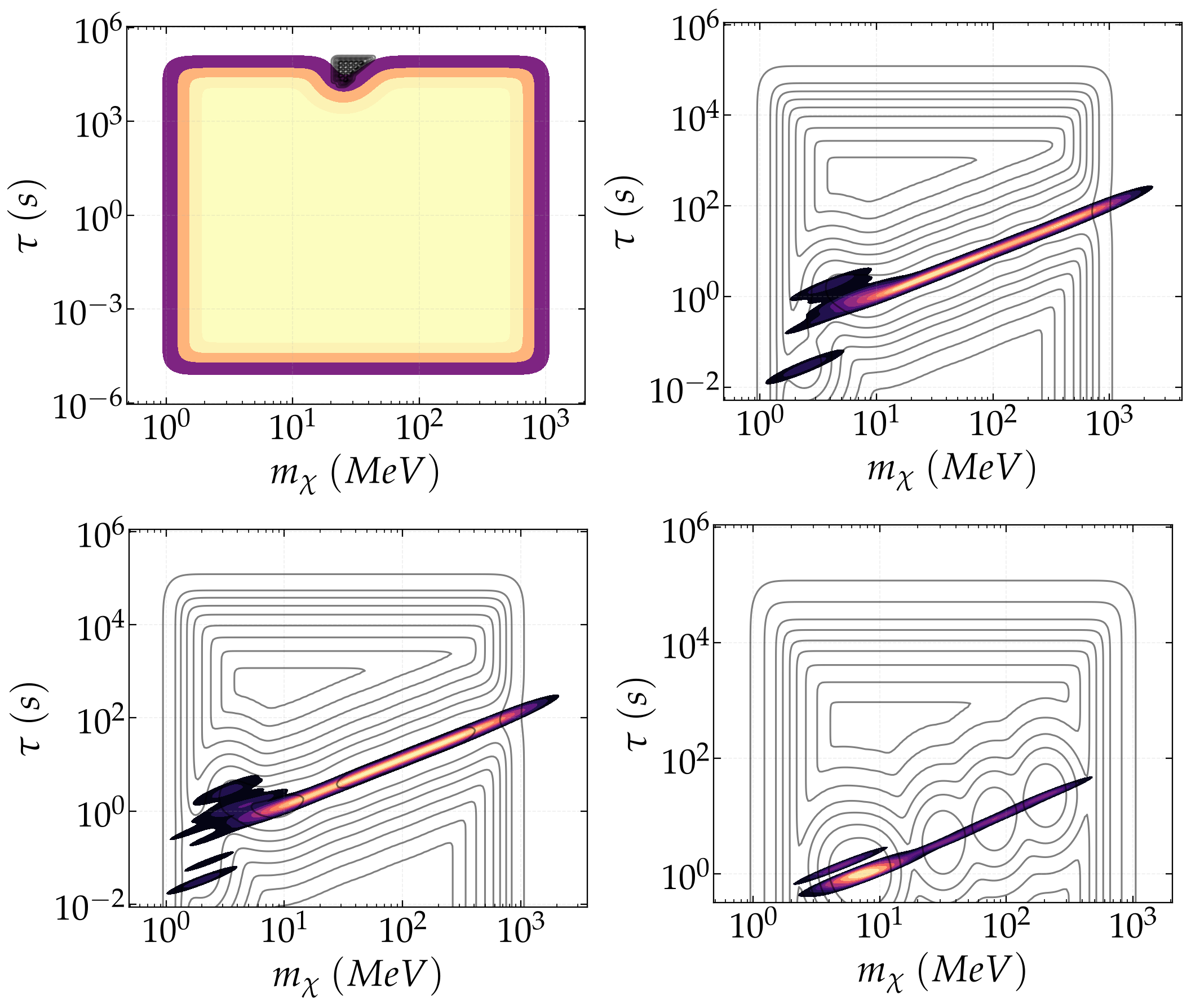}
   \makebox[\linewidth][c]{%
        \makebox[0.48\linewidth][c]{(c)}%
        \hfill
        \makebox[0.48\linewidth][c]{(d)}%
         
    }
\caption{KDE of parameter points satisfying primordial abundance constraints for inital benchmark abundance: (a) $Y_p$, (b) theoretical ${}^7\mathrm{Li/H}$, (c) ${\rm D/H}$, and (d) observational ${}^7\mathrm{Li/H}$. The black contour lines indicate regions that violate the corresponding observational bounds.}
\label{five2}
\end{figure}

\begin{figure}[t]
    \centering
     \makebox[\linewidth][c]{%
        \makebox[0.48\linewidth][c]{(a)}%
        \hfill
        \makebox[0.48\linewidth][c]{(b)}%
         
    }
    \includegraphics[width=\linewidth]{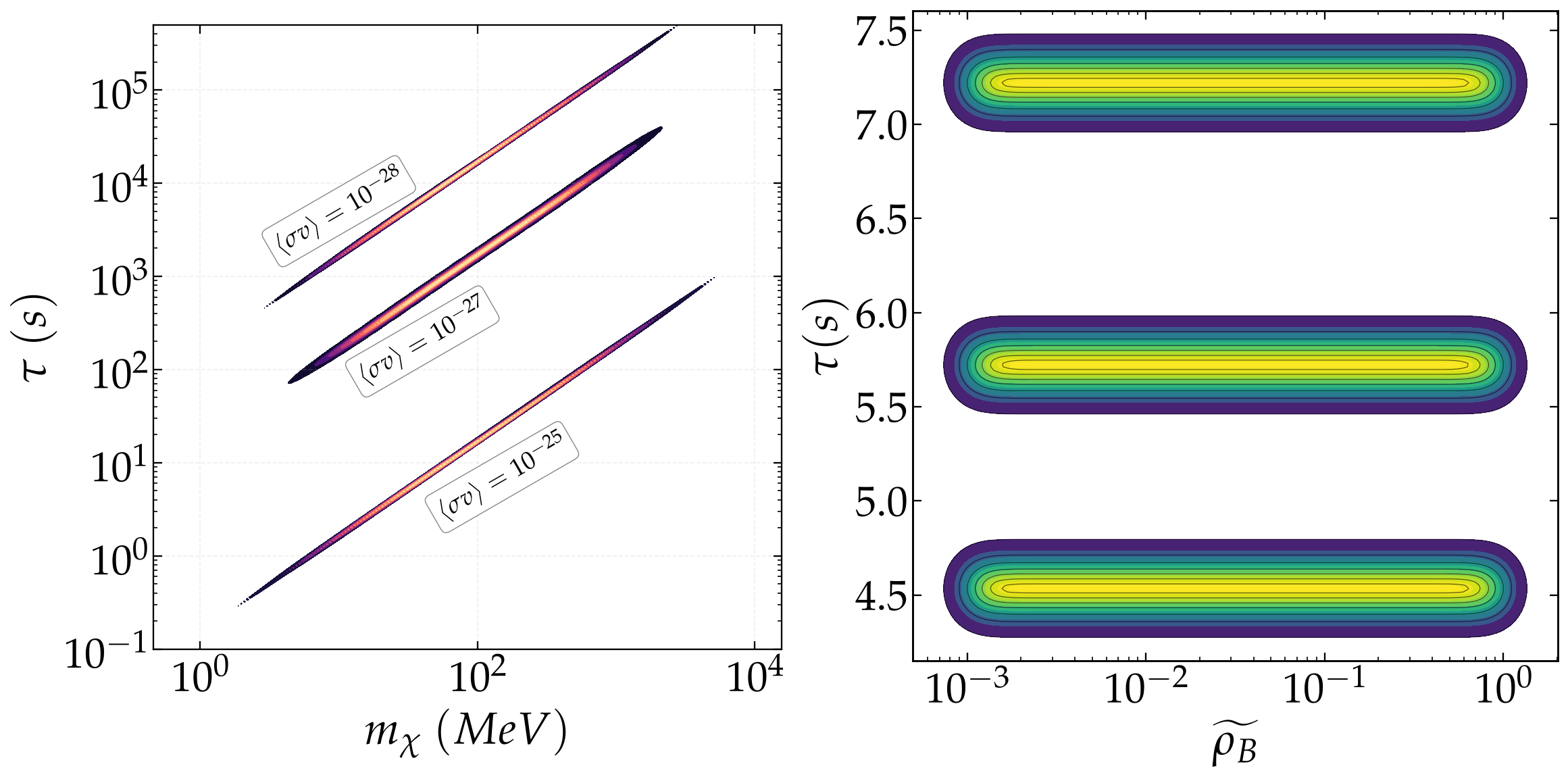}

\caption{KDEs of accepted parameter points for varying cross sections in (a) and $\widetilde{\rho}_{\mathcal{B}}$ in (b).}
\label{six}
\end{figure}
In Fig.~\ref{three}, we present the combined constraints on the parameter space. Regions I, II, III, and IV correspond to parameter points for which  one, two, three, and none of the considered light element abundance constraints are violated, respectively. From region IV, it is evident that the parameter space is predominantly constrained by the deuterium and lithium abundance. Similar to \cite{s1,s2,s3,s4} we find no overlapping region in the parameter space that simultaneously satisfies the observational ${}^7\mathrm{Li/H}$  and the remaining abundance constraints.
\section{Non Thermal Nucleosynthesis}
\label{nth}

\begin{table}[b]
    \caption{SBBN and Benchmark light element abundances\label{tab:benchmarks}}
    \begin{ruledtabular}
    \begin{tabular}{clc}
        Nuclei & $2\sigma$ Constraints & Benchmark   \\
        \colrule
        \noalign{\vskip 3pt}
        $Y_p$ & $0.24493204 \pm 0.008$ \cite{yp}
          & $0.2508 \pm 0.002$ \\
        $\frac{D}{H}$ & $ (2.5 \pm 0.058) \times 10^{-5} $ \cite{dh_li}
          & $ (3\pm 0.1) \times 10^{-5}$  \\
        $\frac{{}^7 Li}{H}$ & $(1.45\pm0.5) \times 10^{-10}$  \cite{dh_li}
          & --  \\
        $\frac{{}^7 Li}{H}_{th}$ & $(5.624 \pm 0.49) \times 10^{-10}$  \cite{li_th}
          & $(4.5 \pm 0.1) \times 10^{-10}$ \\
    \end{tabular}
    \end{ruledtabular}
\end{table}

After the completion of BBN, for times $t \gtrsim 10^{4}\,\mathrm{s}$, residual annihilation or decay of dark matter can produce high energy photons capable of photodisintegrating the light nuclei synthesized during BBN, thereby modifying their primordial abundances \cite{pd1,pd2,pd3,pd4,pd5,pd6}. In primordial BBN, no common parameter space satisfying all observational constraints was found. In this section, we show that the inclusion of photodisintegration opens up a common parameter region consistent with the observations.

To compute the light element abundances after photodisintegration, we modified the publicly available \texttt{ACROPOLIS}  code \cite{pd7}. For the \texttt{cosmo\_file.dat} input, we solved the background cosmological evolution equations, as described in the Supplementary Material, and used the resulting  data as input. For \texttt{abundance\_file.dat}, we give benchmark abundances summarized in Table~\ref{tab:benchmarks} as input. This benchmark satisfies the 2 $\sigma$ bounds on $Y_p$, while D/H is overproduced and the predicted $^7\mathrm{Li}$ abundance is slightly below the theoretically predicted value. We adopt this benchmark because the D/H and ${}^7\mathrm{Li}/\mathrm{H}$ abundances obtained from our preceding BBN calculations and they are compatible with the $2\sigma$ observational constraint on $Y_p$. As we show below, this benchmark allows us to resolve the ${}^7\mathrm{Li}$ problem while remaining consistent with the deuterium and helium abundance. Other benchmark choices either violate the $2\sigma$ constraint on $Y_p$ or lead to excessive deuterium destruction in the parameter regions that successfully address the ${}^7\mathrm{Li}$ problem. This behavior is consistent with the features revealed by the parameter scan discussed in the Supplementary Material.

Figures~\ref{four} and~\ref{five2} show the KDE distributions in the $\tau$--$m_\chi$ parameter space, where $\tau$ denotes the lifetime of the bound state before its decay into SM particles. We fix the self annihilation cross section to $\langle\sigma v\rangle = 10^{-25}\,\mathrm{MeV}^{-2}$, the branching ratios to $BR_{e^\pm}=BR_{2\gamma}=0.5$, and $\widetilde{\rho}_B=10^{-2}$ at $T\simeq 1\,\mathrm{MeV}$.

In both case, the helium abundance remains mostly unaffected by photodisintegration, as the corresponding photodisintegration threshold is relatively high, $E_{\rm th}\simeq 19.81\,\mathrm{MeV}$. For both scenario, there exists a range of $\tau$ values for which the ${}^{7}\mathrm{Li}$ abundance can be reduced to its observationally preferred value over the entire range of $m_\chi$ considered. Since the photodisintegration thresholds of deuterium and ${}^{7}\mathrm{Be}$ are relatively close, $E_{\rm th}^{\rm D}\simeq 2.2\,\mathrm{MeV}$ and $E_{\rm th}^{\rm Be}\simeq 1.5\,\mathrm{MeV}$, the reduction of ${}^{7}\mathrm{Be}$ is  accompanied by the photodisintegration of deuterium and most of remaining ${}^7\mathrm{Be}$ would decay into ${}^7\mathrm{Li}$. Panel (c) of Fig.~\ref{four} clearly demonstrates that, within the SBBN framework, none of the parameter choices yields a D/H abundance compatible with observations.

In contrast, Benchmark  starts with an initially enhanced deuterium abundance and underproduced lithium. Photodisintegration can therefore simultaneously reduce both $\frac{D}{H}$ and $\frac{^7 Li}{H}$ abundances to their observationally preferred values for a common range of $\tau$ and $m_\chi$, as illustrated in panels~(b) and~(c) of Fig.~\ref{five2}. While existing photodisintegration solutions to the \({}^7\mathrm{Li}\) problem require \(m_\chi \sim 3\text{--}4\,\mathrm{MeV}\) \cite{de1,de2}, our framework resolves it across the entire dark matter mass range. As shown in panel (d), Benchmark admits a common region of parameter space that satisfies the observational constraints on $D/H$, $Y_p$, and ${}^{7}\mathrm{Li}/H$, thereby providing a simultaneous resolution of the ${}^{7}\mathrm{Li}$ abundance discrepancy.
In Figure \ref{six} panel (a), we fix $\widetilde{\rho}_B = 10^{-2}$, while in the other panel we fix $m_\chi = 100\,\mathrm{MeV}$ and $\langle\sigma v\rangle = 10^{-25}\,\mathrm{MeV}^{-2}$. We find that the abundance remains largely insensitive to variations in $\widetilde{\rho}_B$. Furthermore, as the annihilation cross section decreases, bound states with progressively longer lifetimes are required to reproduce the observed abundances.

\section{Conclusion}
\label{con}
Our thermal nucleosynthesis analysis shows that the framework is compatible with the theoretically predicted light element abundances over distinct regions of the parameter space. For $\widetilde{\rho}_{\mathcal B}<10^{-2}$, allowed lifetimes are in the range $10^{4}s<\tau<10^{8}s$, while for $\widetilde{\rho}_{\mathcal B}>10^{-2}$, the allowed range shifts to $10^{-4}s<\tau<10^{0}s$. Although $10^4<\tau<10^{8}s$ can alleviate the cosmological ${}^7$Li problem, this region remains in tension with the abundances of the other light elements.

We find that this tension can be resolved by incorporating non thermal nucleosynthesis,using the abundances predicted by thermal nucleosynthesis as the initial conditions. In Benchmark, for all values of $m_\chi$ considered, bound states with lifetimes in the range $10^0\mathrm{s}\lesssim\tau\lesssim10^2 \mathrm{s}$ can simultaneously alleviate the ${}^7$Li problem while satisfying the constraints from the other light element abundances at 2 $\sigma$ level. The results are largely insensitive to variations in $\widetilde{\rho}_{\mathcal B}$. Moreover, as the annihilation cross section decreases,  longer lived bound states are required to resolve the ${}^7$Li discrepancy without violating the remaining nucleosynthesis constraints.
\onecolumngrid

\appendix
\section{Supplementary Material}
\subsection{Background cosmology}

\label{sec:param}

The presence of dark matter ($\chi$),dark mediator ($\Phi$), and bound state ( $\mathcal{B}$) modifies the standard cosmology and alters many quantities such as temperature, entropy and expansion rate of universe. The of dark sector gives a addidtional contibution to enrgy density of universe, the total energy density of the Universe, which can be written as
\begin{equation}
  \label{eq:nt}
  \rho_{\mathrm{tot}} = \rho_{\mathrm{SM}} + \rho_{\mathrm{DS}},
\end{equation}
where $\rho_{\mathrm{SM}}$ and $\rho_{\mathrm{DS}}$ denote the energy densities of the Standard Model  sector and the dark sector, respectively.

The dark sector energy density receives contributions from free dark matter ($\rho_\chi$), bound state ($\rho_{\mathcal{B}}$), and dark mediator ($\rho_{\Phi}$). The evolution equations for the energy densities of these three dark sector components are given by
\begin{equation}
  \label{eq:ex}
  \dot{\rho_\chi} + 3 H \rho_\chi 
  = -\frac{\rho_\chi^2 \langle \sigma v_{BSF}\rangle}{m_\chi} 
    - \frac{\rho_\chi^2 \langle \sigma v_{DR}\rangle}{m_\chi} 
    - \frac{\rho_\chi^2 \langle \sigma v_{SM}\rangle}{m_\chi} 
    + \rho_\mathcal{B} \rho_{\Phi} \Gamma_{Ion},
\end{equation}
\begin{equation}
  \label{eq:eb}
  \dot{\rho_\mathcal{B}} + 3 H \rho_\mathcal{B} 
  = \frac{\rho_\chi^2 \langle \sigma v_{BSF}\rangle}{m_\chi} 
    - \rho_\mathcal{B} \Gamma_{DR} 
    - \rho_\mathcal{B} \Gamma_{SM}
    - \rho_\mathcal{B} \rho_{\Phi} \Gamma_{Ion},
\end{equation}
\begin{equation}
  \label{eq:ea}
  \dot{\rho_\Phi} + 4 H \rho_\Phi
  = \frac{\rho_\chi^2 \langle \sigma v_{BSF}\rangle}{m_\chi} 
    + \rho_\mathcal{B} \Gamma_{DR} 
    + \frac{\rho_\chi^2 \langle \sigma v_{DR}\rangle}{m_\chi} 
    - \rho_\mathcal{B} \rho_{\Phi} \Gamma_{Ion}.
\end{equation}

In the above equations, $\rho_i = n_i m_i$. The quantity $\Gamma_{DR}$ denotes the decay width of bound state into a pair of dark mediators, while $\Gamma_{SM}$ corresponds to the decay into a pair of SM particles. The terms $\langle \sigma v_{DR} \rangle$ and $\langle \sigma v_{SM} \rangle$ represent the thermally averaged annihilation cross sections of free dark matter into dark radiation and SM particles, respectively.

The total dark sector energy density evolves according to the sum of Eqs.~\eqref{eq:ex}–\eqref{eq:ea}, which yields
\begin{equation}
  \dot{\rho_{DS}} + 3H(\rho_{DS}+ P_{DS}) 
  = \frac{\rho_\chi^2 \langle \sigma v_{BSF}\rangle}{m_\chi} 
    - \rho_\mathcal{B} \rho_{\Phi} \Gamma_{Ion}
    - \rho_\mathcal{B} \Gamma_{SM} 
    - \frac{\rho_\chi^2 \langle \sigma v_{SM}\rangle}{m_\chi}.
\end{equation}

Since $\dot{\rho}_{\mathrm{tot}}+3H(\rho_{tot}+P_{tot}) = 0$, using Eq.~\eqref{eq:nt} we obtain
\begin{equation}
  \label{eq:eqrhosm}
  \dot{\rho_{\mathrm{SM}}} + 3H(\rho_{\mathrm{SM}}+ P_{\mathrm{SM}}) 
  = -\frac{\rho_\chi^2 \langle \sigma v_{BSF}\rangle}{m_\chi} 
    + \rho_\mathcal{B} \rho_{\Phi} \Gamma_{Ion} 
    + \rho_\mathcal{B} \Gamma_{SM} 
    + \frac{\rho_\chi^2 \langle \sigma v_{SM}\rangle}{m_\chi}.
\end{equation}
\begin{figure}
  \centering
  \includegraphics[width=0.75\textwidth]{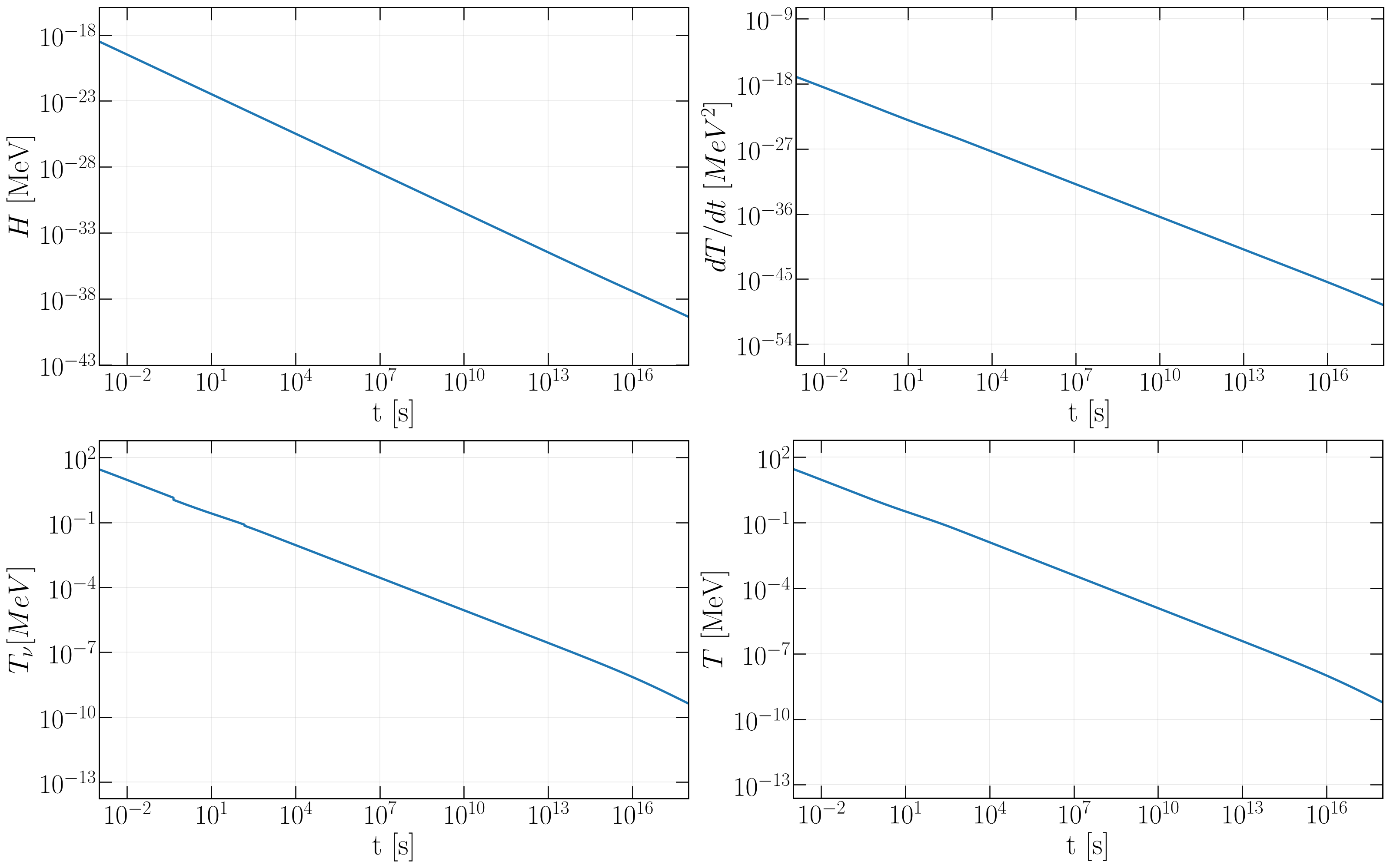}
  \caption{Evolution of the background parameters, including the temperature of the thermal bath, the neutrino temperature, and the Hubble expansion rate.}
  \label{param}
\end{figure}

 In this work, we assume that dark matter annihilates or decays exclusively into the electromagnetic (EM) sector of the Standard Model. Annihilation or decay into neutrinos is beyond the scope of this study. Under this assumption, the SM energy density and pressure before neutrino decoupling ($t < t_{\nu\text{-dec}}$) are given by Eqs.~\eqref{eq:rsm} and \eqref{eq:psm},
\begin{equation}
  \label{eq:rsm}
  \rho_{\mathrm{SM}}=
  \frac{\pi^2}{30} 
  g_e(T(t<t_{\nu\text{-dec}})) 
  T^4(t<t_{\nu\text{-dec}}),
\end{equation}
\begin{equation}
  \label{eq:psm}
  P_{\mathrm{SM}}=
  \frac{\pi^2}{90} 
  g_p(T(t<t_{\nu\text{-dec}})) 
  T^4(t<t_{\nu\text{-dec}}).
\end{equation}

Substituting Eqs.~\eqref{eq:rsm} and \eqref{eq:psm} into Eq.~\eqref{eq:eqrhosm}, we obtain Eq.~\eqref{eq:t1}, which governs the evolution of the SM bath temperature prior to neutrino decoupling:
\begin{equation}
  \label{eq:t1}
  \dot{T}(t<t_{\nu\text{-dec}}) 
  = \frac{30 C[f]}{\pi^2 T^3 \beta}
    - \frac{3 H T \alpha}{\beta}.
\end{equation}
Here,
\[
C[f]= -\frac{\rho_\chi^2 \langle \sigma v_{BSF}\rangle}{m_\chi} 
+ \rho_\mathcal{B} \rho_{A^\mu} \langle \sigma v_{Ion}\rangle 
+ \rho_\mathcal{B} \Gamma_{SM} 
+ \frac{\rho_\chi^2 \langle \sigma v_{SM}\rangle}{m_\chi},
\]
is the collision term,
\[
\alpha = \frac{g_p(T)}{3}+g_e(T), 
\qquad 
\beta=4 g_e(T)+T \frac{d g_e(T)}{dT}.
\]

For times after neutrino decoupling ($t>t_{\nu\text{-dec}}$), neutrinos are no longer in thermal equilibrium with the SM bath. The energy and pressure densities are therefore modified as in Eqs.~\eqref{eq:rhond} and \eqref{eq:pnd},
\begin{equation}
  \label{eq:rhond}
  \rho_{\mathrm{SM}}=
  \frac{\pi^2}{30} 
  g_e'(T(t>t_{\nu\text{-dec}})) 
  T^4(t>t_{\nu\text{-dec}}+\frac{42}{8}T_\nu^4),
\end{equation}
\begin{equation}
  \label{eq:pnd}
  P_{\mathrm{SM}}=
  \frac{\pi^2}{90} 
  g_p'(T(t>t_{\nu\text{-dec}})) 
  T^4(t>t_{\nu\text{-dec}}+\frac{42}{8}T_\nu^4).
\end{equation}

Here, $g_e'$ and $g_p'$ denote the effective degrees of freedom after neutrino decoupling. Substituting Eqs.~\eqref{eq:rhond} and \eqref{eq:pnd} into Eq.~\eqref{eq:eqrhosm}, we obtain Eq.~\eqref{eq:t2}, which determines the evolution of the SM bath temperature after neutrino decoupling:
\begin{equation}
  \label{eq:t2}
  \dot{T}(t>t_{\nu\text{-dec}}) 
  = \frac{30 C[f]}{\pi^2 T^3 \beta}
    - \frac{3 H T \alpha}{\beta}.
\end{equation}
In this case,
\[
\alpha = \frac{g_p'(T)}{3}+g_e'(T), 
\qquad 
\beta=4 g_e'(T)+T \frac{d g_e'(T)}{dT}.
\]

The entropy  of the Universe is also modified by the introduction of new particles such as $\chi$, $\mathcal{B}$, and $\Phi$. Although the dark mediator is not directly coupled to the SM thermal bath, dark matter is assumed to be in thermal equilibrium with both the SM and the dark mediator. Consequently, the dark mediator can also attain thermal equilibrium with the SM bath through its coupling to dark matter. Conisdering all  these, the entropy density could be defined in terms of $\rho_{SM}$ as,
\begin{equation}
S(T) = \frac{4}{3}
\left(\frac{\pi^2 g_{*}(T)}{30}\right)^{1/4}
\rho_{\rm SM}^{3/4}.
\end{equation}

Figure \ref{param} shows the numerical evolution of the background quantities relevant for the non thermal nucleosynthesis calculation, namely the temperature of the thermal bath, the neutrino temperature, the temperature evolution rate $dT/dt$, and the Hubble expansion rate. As the Universe evolves with time, all four quantities exhibit the expected redshift behavior associated with the cosmological expansion.  We use these numerical data as inputs to the \texttt{cosmo\_file.dat} file of the \texttt{Acropolis} code, which subsequently uses the evolving cosmological background to compute the effects of non thermal processes on the primordial nucleosynthesis and the resulting light element abundances.

\section{Thermal Nucleosynthesis } 
Lighter dark matter ($m_\chi < 30~\text{MeV}$) freezes out during BBN, with the bound state assumed to be stable 
through nucleosynthesis and their energy density would be given by,
\begin{figure}
  \centering
  \includegraphics[width=\textwidth]{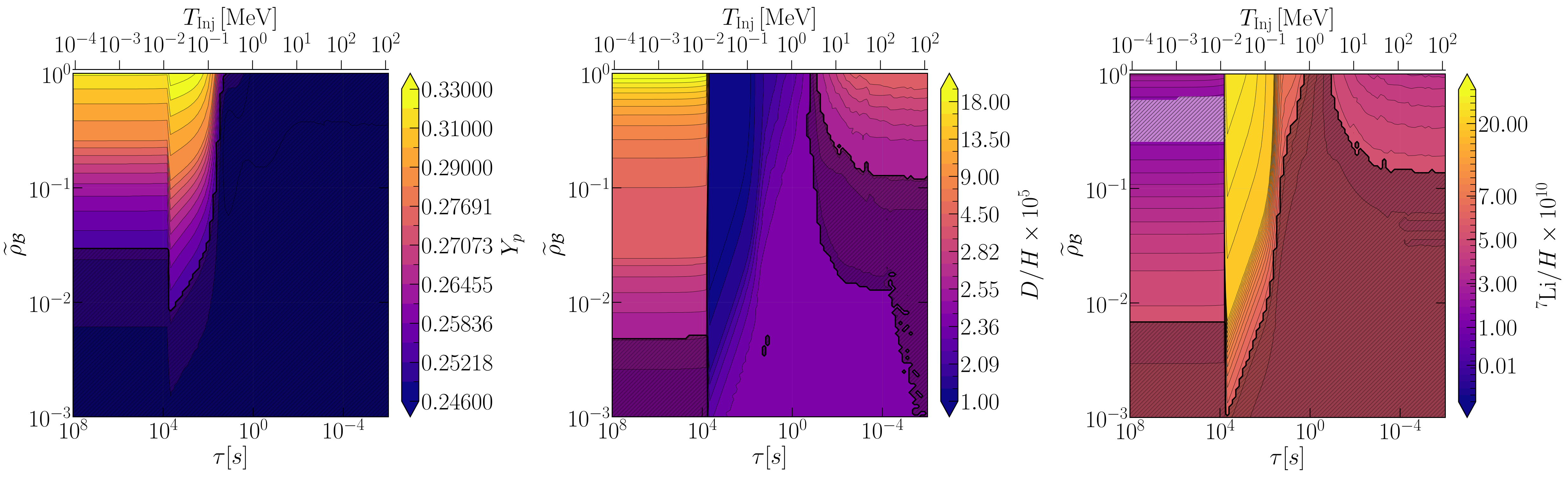}
\caption{Exclusion plots for the primordial abundances of $Y_p$, D/H, and $^7$Li after the completion of primordial nucleosynthesis. The shaded regions indicate the parameter space allowed by the theoretical predictions, while the white-shaded region corresponds to the parameter space satisfying the experimentally inferred bounds on the primordial $^7$Li abundance.}
\label{thn}
\end{figure}

\begin{equation}
    \rho_\chi=g_\chi m_\chi^4 \sum_{1}^{\infty} cosh(\frac{n \mu_\chi}{T_\chi}) M(n ~x)
   \end{equation}
   \begin{equation}
    \rho_B=g_B m_B^4 \sum_{1}^{\infty} cosh(\frac{n \mu_B}{T_B}) M(n ~x)
   \end{equation}
   \begin{equation}
    \rho_\Phi=k_{dr} \rho_\gamma (T=1 MeV) T^4
   \end{equation}
   where $M(x)=\frac{1}{x} \big(\frac{3 K_3(x)}{4}+\frac{K_1(x)}{4}\big)$ and $k_{dr}$ is $\frac{\rho_\Phi}{\rho_\gamma}$ at $T=1~MeV$.
But for $m_\chi > 30~\mathrm{MeV}$, the bound states decay out of equilibrium, and the energy densities evolve as:
\begin{equation}
  \dot{\rho_\mathcal{B}} + 3 H \rho_\mathcal{B} 
  =  - \rho_\mathcal{B} \Gamma_{DR} 
    - \rho_\mathcal{B} \Gamma_{SM} - \rho_\Phi \rho_\mathcal{B} \Gamma_{Ion}
\end{equation}
\begin{equation}
  \dot{\rho_\Phi} + 4 H \rho_\Phi
  =  + \rho_\mathcal{B} \Gamma_{DR} - \rho_\Phi \rho_\mathcal{B} \Gamma_{Ion}
\end{equation}
However, we find that variations in $\langle \sigma v\rangle_{\mathrm{BSF}}$, which modify $\Gamma_{\mathrm{Ion}}$, do not significantly affect the results, indicating that primordial BBN is insensitive to these variations.
The presence of the additional species modifies the baryon-to-photon ratio through their contributions to the total entropy density. In particular, the modified ratio can be expressed as

$$
\eta = \eta_0\left[1+\frac{s_e(T)+s_\chi(T)+s_\Phi(T)+s_B(T)}
{s_\gamma(T)}\right],
$$

where $\eta_0$ denotes the standard baryon-to-photon ratio, while $s_\chi(T)$, $s_\Phi(T)$, and $s_B(T)$ represent the entropy density contributions from the corresponding species.
In Fig.~\ref{thn}, we present the contour plots of the primordial light-element abundances, together with the corresponding observational constraints listed in Benchmark I.

\section{Non thermal nucleosynthesis}
Once the universe cools sufficiently below the temperature where $e^\pm$ pair creation becomes ineffective, high energy photons can induce the photodissociation of light nuclei. Specifically, the  temperature and photon energy thresholds for nuclear disintegration are $T < 0.6\text{ keV}$ and $E_\gamma > 19.81\text{ MeV}$ for helium, $T < 5.34\text{ keV}$ and $E_\gamma > 2.2\text{ MeV}$ for deuterium , $T < 4.81\text{ keV}$ and $E_\gamma > 2.47\text{ MeV}$ for lithium-7 , and $T < 7.48\text{ keV}$ and $E_\gamma > 1.59\text{ MeV}$ for beryllium-7.
Following Ref.~\cite{pd7}, when the photodissociation rate exceeds the Hubble expansion rate ($\Gamma_p > H$), the Boltzmann evolution equation can be expressed as:
\begin{equation}
    \frac{\partial f_\gamma(E)}{\partial t} \approx S_\gamma(E) - \Gamma_\gamma f_\gamma(E) + \sum_{x} \int_{E}^{\infty} \mathrm{d}E' \, \mathrm{d}\Gamma_{(x,\gamma)}(E,E') f_x(E'),
\end{equation}
where $S_\gamma(E)$ is the photon source term at energy $E$, $\Gamma_\gamma$ denotes the photon interaction rate, and $\mathrm{d}\Gamma_{(x,\gamma)}$ represents the differential interaction rate for a particle species $x$ with energy $E'$ to produce photons of energy $E$.
Having solved for $f_\gamma$, the yield evolution of light nuclei can be determined via the following  equation:
\begin{equation}
    \dot{Y}_N = \sum_j Y_j \int_{0}^{\infty} \mathrm{d}E \, f_\gamma(t,E) \sigma_{j\gamma \to N} - Y_N \sum_{j'} \int_{0}^{\infty} \mathrm{d}E \, f_\gamma(t,E) \sigma_{N\gamma \to j'},
\end{equation}
where $Y$ denotes the yields of the light nuclei, and $\sigma$ represents the respective cross sections for the reactions.
To evaluate these equations, two key quantities are required: $E_0$, representing the energy of particles produced by dark matter annihilation or decay (such that the resulting photons have energies $E < E_0$), and the total source term, which can be expressed as $S = S_0 + S_{\text{FSR}}$, where $S_0$ is the monochromatic source and $S_{\text{FSR}}$ accounts for final-state radiation.

For the decay of a bound state, $E_0 = m_B / 2$, whereas for the annihilation of free dark matter, $E_0 = m_\chi$. The monochromatic source terms for the decay scenario are given by:
\begin{align}
    S_0^\gamma &= \text{BR}_\gamma \frac{2 n_B}{\tau_{\text{SM}}}, \\
    S_0^{e^\pm} &= \text{BR}_{e^\pm} \frac{2 n_B}{\tau_{\text{SM}}},
\end{align}
where the number density $n_B$ evolves as:
\begin{equation}
    n_B(t) = n_B(t_0) \left(\frac{R(t_0)}{R(t)}\right)^3 \exp\left(-\frac{t}{\tau_{\text{SM}}} - \frac{t}{\tau_{\text{DR}}}\right).
\end{equation}
Here, $\tau_{\text{SM}}$ is the lifetime for decay into Standard Model particles, and $\tau_{\text{DR}}$ is the lifetime for decay into dark radiation. 
\begin{figure}[t]
    \centering
     \makebox[\linewidth][c]{%
        \makebox[0.48\linewidth][c]{(a)}%
        \hfill
        \makebox[0.48\linewidth][c]{(b)}%
         
    }
    \includegraphics[width=0.75\linewidth]{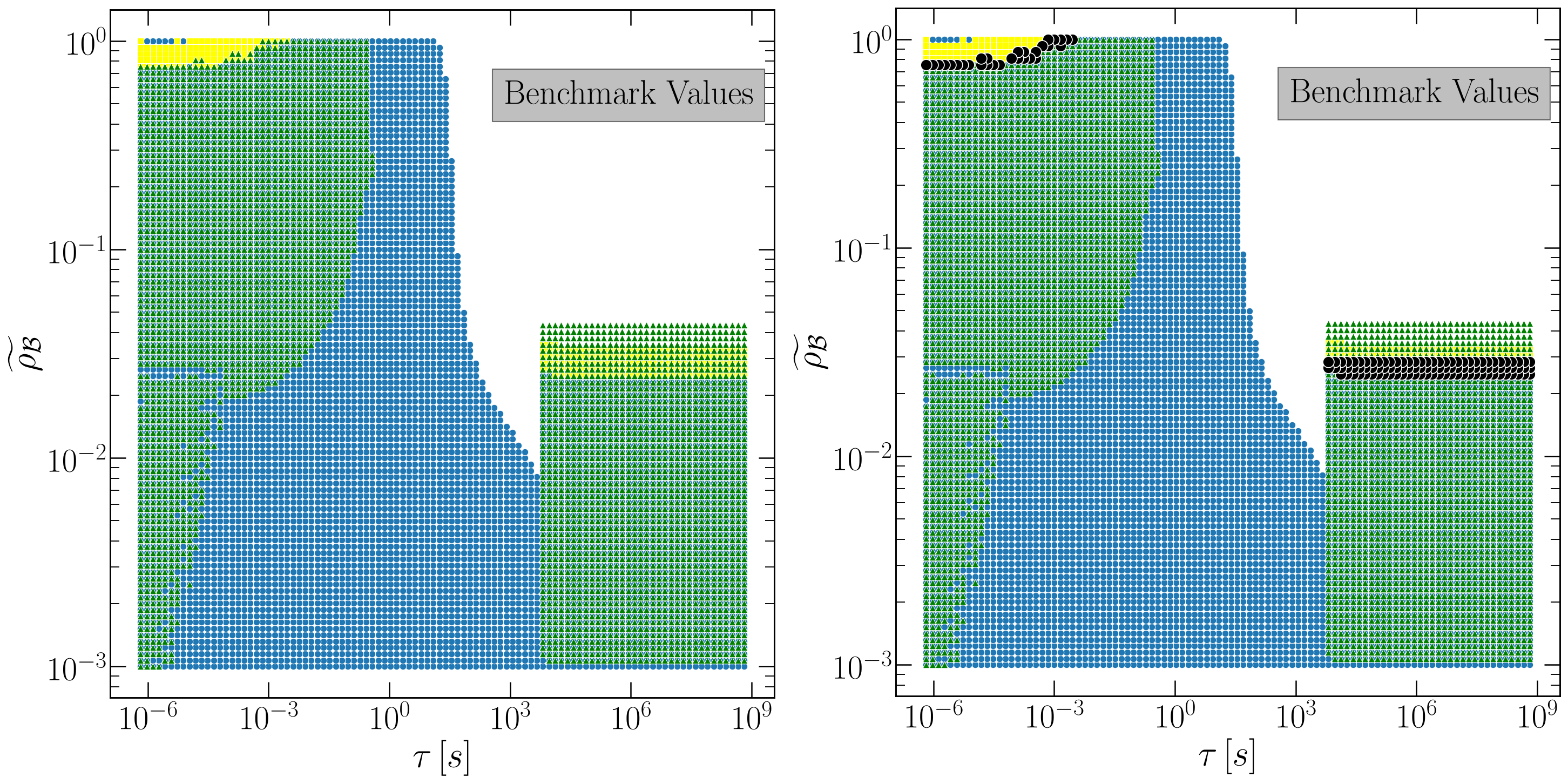}
     \makebox[\linewidth][c]{%
        \makebox[0.48\linewidth][c]{(c)}%
        \hfill
        \makebox[0.48\linewidth][c]{(d)}%
         
    }
\caption{Combined allowed parameter space for Benchmark. The blue, yellow, and green regions represent parameter points satisfying the observational constraints on $Y_p$, ${\rm D/H}$, and ${}^7\mathrm{Li/H}$, respectively. The black region denotes the common parameter space satisfying all three abundance constraints simultaneously.}
\label{five3}
\end{figure}
Similarly, the monochromatic source terms for the annihilation of free dark matter are expressed as:
\begin{align}
    S_0^\gamma &= \text{BR}_\gamma \langle \sigma v \rangle n_\chi^2, \\
    S_0^{e^\pm} &= \text{BR}_{e^\pm} \langle \sigma v \rangle n_\chi^2,
\end{align}
where $n_\chi$ is the dark matter number density. In general, the thermally averaged cross section is given by $\langle \sigma v \rangle = a + b \langle v^2 \rangle$. However, since we are considering purely s-wave processes, we have $\langle \sigma v \rangle \approx a$.

Finally, the source terms for final-state radiation (FSR) in both cases are given by:
\begin{align}
    S_{\text{FSR}}^{e^\pm} &= 0, \\
    S_{\text{FSR}}^\gamma &= \frac{S_0^{e^\pm}}{E_0} \frac{\alpha \left[1 + (1-x)^2\right]}{\pi x} \ln\left(\frac{4 E_0^2 (1-x)}{m_e^2}\right) \Theta\left(1 - \frac{m_e^2}{4 E_0^2} - x\right),
\end{align}
where $x = E / E_0$ and $\alpha$ is the fine-structure constant.
In Fig.~\ref{five3}, we present the parameter spaces satisfying the light element abundance constraints for Benchmarks we have chosen and it  is chosen such that the  ${\rm D/H}$ and ${}^7\mathrm{Li/H}$ abundances are consistent with the observational $2\sigma$ bound on $Y_p$ and alleviate the ${}^7\mathrm{Li}$ tension. We adopt an enhanced ${\rm D/H}$ abundance for this Benchmark  because photodisintegration, which reduces the ${}^7\mathrm{Li}$ abundance, can also deplete deuterium owing to the relatively close energy thresholds for their destruction. By selecting a higher initial ${\rm D/H}$ abundance and the corresponding ${}^7\mathrm{Li/H}$ and $Y_p$ values, we aim to ensure that the abundances remain consistent with observations after photodisintegration. For ${\rm D/H}$ values higher than that of Benchmark , we find no common parameter space in which the ${}^7\mathrm{Li/H}$ and ${\rm D/H}$ abundances satisfy their observational constraints simultaneously with the $2\sigma$ bound on $Y_p$. Although lower initial ${\rm D/H}$ values can yield a common allowed parameter space, the subsequent depletion of deuterium during photodisintegration leads to a final abundance below the observed range.

\bibliography{ref}

\end{document}